\documentclass[12pt]{iopart}
\usepackage{iopams}  
\usepackage{epsfig}
\usepackage{graphicx}
\usepackage{epstopdf}

\def\slashchar#1{\setbox0=\hbox{$#1$}     		
   \dimen0=\wd0                                 	
   \setbox1=\hbox{/} \dimen1=\wd1               	
   \ifdim\dimen0>\dimen1                        	
      \rlap{\hbox to \dimen0{\hfil/\hfil}}      	
      #1                                        	
   \else                                        	
      \rlap{\hbox to \dimen1{\hfil$#1$\hfil}}   	
      /                                         	
   \fi}

\renewcommand{\vec}{\boldsymbol }

\newcommand{\qw}{Q_\mathrm{w}}
\newcommand{\ncs}{N_\mathrm{CS}}

\begin{document}

\title{Implications of $\mathcal{CP}$-violating
transitions in hot quark matter on heavy ion collisions}

\author{Harmen J.\ Warringa}

\address{Department of Physics,
Bldg.\ 510A, 
Brookhaven National Laboratory,\\
Upton NY 11973, USA}
\ead{warringa@quark.phy.bnl.gov}

\begin{abstract}
  Quantum Chromodynamics (QCD) predicts that topological charge
  changing transitions will take place in hot quark matter.  Such
  transitions induce $\mathcal{P}$- and $\mathcal{CP}$-violating
  effects.  We will show that in the presence of a magnetic field
  these transitions can separate quarks according to their electric
  charge along the direction of the magnetic field. This is
  the so-called Chiral Magnetic Effect. We will argue that
  it might be possible to observe the Chiral Magnetic Effect
  in heavy ion collisions. 
\end{abstract}

\section{Introduction}
Although the ground state of QCD at zero temperature cannot break the
parity ($\mathcal{P}$) and charge-parity ($\mathcal{CP}$) symmetries
spontaneously \cite{VW}, this is not necessarily true anymore at
finite temperatures and/or chemical potentials \cite{VWth}.  Hence the
possibility of spontaneous breakdown of discrete symmetries of space
and time arises \cite{DL}, which could be realized in heavy ion
collisions \cite{Morley}.  In Ref.~\cite{KPT} it was argued that
during the chiral symmetry breaking phase transition the matter
produced in a heavy ion collision may cool to a meta-stable
vacuum. This meta-stable vacuum can effectively be described by QCD
with a finite $\theta$ angle. The decay of such a meta-stable vacuum
could give rise to all kinds of interesting $\mathcal{P}$- and
$\mathcal{CP}$-violating behavior \cite{decay}, which might be
detected in experiment \cite{Voloshin-finch} by using suitable
observables \cite{KP}.

Kharzeev has pointed out that if $\mathcal{P}$- and
$\mathcal{CP}$-violating processes are taking place in the quark
matter produced in heavy ion collisions, this will lead to separation
of electric charge along the direction of angular momentum of the
collision \cite{Kharzeev}.  This in some sense similar to an electric
dipole moment, but now the direction of the dipole moment is expected
to fluctuate from event-to-event.  Voloshin has shown that this effect
can be studied experimentally by analyzing correlations between
charged particles and the reaction plane
\cite{Voloshin2004}. Preliminary data of the STAR collaboration is
presented in Ref.~\cite{IVS}. The scenario of Kharzeev \cite{Kharzeev}
was worked out in more detail in Refs.~\cite{KZ, KMW} and
called the Chiral Magnetic Effect: topological charge changing
transitions induce chirality which leads to separation of charge along
the direction of the magnetic field. We will study this effect in
detail in this article.

\section{Generating chirality} 
We will assume that the quarks are massless. This is
expected to be a reasonable approximation in the quark gluon plasma
phase where the typical momenta of the quarks are much larger than
their masses.  In that case right-handed quarks and anti-quarks have
spin and momentum in the same direction, while left-handed ones have
them in the opposite direction. The projection of spin on the momentum
is called helicity. Since we have assumed that the quark are massless
this is the same as chirality.

Massless quarks can change their chirality by interacting with gluons.
In QCD there is an exact relation, the so-called axial Ward-Identity which
relates the chirality change to the properties of the gluon fields. This
identity arises from the axial anomaly \cite{ABJ} and reads
\begin{equation}
(N_L - N_R) ({t = \infty})
-
(N_L - N_R)({t = -\infty})
= 2 \qw.
\label{eq:axialwi}
\end{equation}
In this equation $N_{L,R}$ stands for the total number of
left/right-handed quarks plus anti-quarks of a particular flavor in
the background of a certain gluon field. The change of chirality of
the quarks is equal to twice the winding number $\qw$ of the gluon fields.
This winding number can be computed as follows
\begin{equation}
\qw = \frac{g^2}{32\pi^2} \int \mathrm{d}^4 x\, 
F_{\mu \nu}^a \tilde F^{\mu\nu}_a.
\end{equation}
Here $g$ denotes the QCD coupling constant 
with generators normalized as $\mathrm{tr}\, t_a t_b = \delta_{ab} / 2$.
The gluonic field tensor and its dual are respectively
$F_{\mu\nu}^a$ and $\tilde F^a_{\mu\nu} = 
\epsilon_{\mu\nu}^{\phantom{\mu\nu}\rho\sigma} F^a_{\rho \sigma} / 2$.
For gluon fields which go to a vacuum solution at $t=\pm \infty$
the winding number $\qw$ is an integer as we will argue in the next section.

In perturbative QCD it is impossible to change the chirality of
massless quarks. This can be easily inferred from
Eq.~(\ref{eq:axialwi}) since in perturbative QCD one only takes into
account gluon fields with $\qw = 0$.  This is a good approximation at
very high energies where the strong coupling constant $\alpha_S$ is
small, but as we will see in more detail in the next section, in the
non-perturbative regime the gluon fields which a nonzero winding
number can give a significant contribution to physical quantities.
One famous example is the mass of the $\eta'$ meson.
A very important conclusion we can draw now from
Eq.~(\ref{eq:axialwi}) is that chirality change and hence
$\mathcal{P}$- and $\mathcal{CP}$-violation are directly linked to the
topology of the gluon fields. 
If one observes a difference between the
number of left- and right-handed fermions, this immediately tells us
that $\mathcal{P}$ and $\mathcal{CP}$ are violated on an
event-by-event basis. Moreover it will be a direct proof 
the existence of topologically non-trivial gluon fields.

\section{Generating winding number}
\begin{figure}[htb]
\begin{center}
\includegraphics[width=10cm]{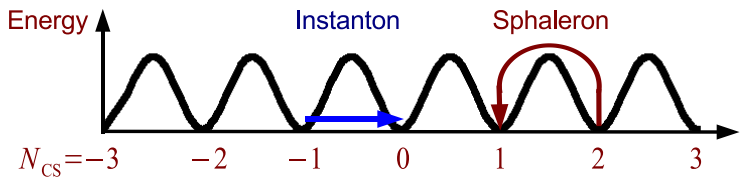}
\end{center}
\caption{Illustration of the vacuum structure of an $\mathrm{SU}(3)$
  Yang-Mills theory.  The different vacua can be characterized by a
  Chern-Simons number $\ncs$, which is an integer. An instanton
  interpolates from one vacuum to another by tunneling, whereas
  a sphaleron
  does this via hopping over this barrier.}
\label{fig:vacuum}
\end{figure}

Now how can gluon fields
wind, and why?  In order to answer that question let
us for a moment forget about the quarks and have look to the vacuum
structure of a pure $\mathrm{SU}(3)$ gauge theory. In the vacuum the
energy density of the gauge fields is minimal, which implies that the
gauge fields are static and have to be a pure gauge. In the temporal
gauge ($A_0 = 0$) one then finds that $A_i(\vec x) = i g^{-1}
U(\vec x) \partial_i U^{\dagger}(\vec x)$, where $U(\vec x)$ is an
element of the gauge group $\mathrm{SU}(3)$.  It is now possible to
assign to each classical vacuum a topological invariant, the
Chern-Simons number $\ncs$ \cite{chern}. For the
vacua this number is an integer and
given by
\begin{equation}
 \ncs = \frac{1}{24\pi^2}
\int \mathrm{d^3}x 
\,
\epsilon^{ijk} \mathrm{tr}
\left[
(U^\dagger \partial_i U)
(U^\dagger \partial_j U)
(U^\dagger \partial_k U)
\right].
\end{equation}
We have illustrated this in Fig.~\ref{fig:vacuum}. The different vacua
are separated by an energy barrier of order $\Lambda_{\mathrm{QCD}}$.
A gauge field configuration with a certain winding number $\qw$
interpolates between two classical vacua. One can show that
\begin{equation}
\qw = \ncs(t=\infty) - \ncs(t=-\infty).
\end{equation}

At zero temperature the only possibility to go from one vacuum to
another is by tunneling through the potential barrier. The gauge field
configuration responsible for such a tunneling process is called an
instanton \cite{hooft}. The rate is exponentially suppressed, even at
finite temperature \cite{pisarski}.

At finite temperature another possibility arises, which is hopping
over the barier. The gauge field configuration which just hops over
the barier is called a sphaleron \cite{sphaleron}.  The rate is not
exponentially suppressed. The rate $\Gamma$ has been computed on the
lattice for $\mathrm{SU}(2)$ Yang-Mills at high temperatures
\cite{moore}. Extrapolating this result to $\mathrm{SU}(3)$ gives
\begin{equation}
\Gamma = \frac{\mathrm{d}N}{\mathrm{d^3 x}\,\mathrm{d t}} 
\approx 386 \alpha_S^5 T^4.
\end{equation}
If the density of quarks is small, the rate will not be changed much
when massless quarks are taken into account.  Hence QCD predicts that
in a thermalized quark gluon plasma several sphaleron transitions per
$\mathrm{fm}^3$ per $\mathrm{fm}/c$ will take place.

In our estimations for the amount of $\mathcal{P}$ and $\mathcal{CP}$
violation in heavy ion collisions we will use the sphaleron rate.
This is correct if a quark gluon plasma has been
formed.  But as was shown in \cite{glasma} the initial state of the
matter produced in a heavy ion collision, the so-called glasma, is
also capable of generating differences in the Chern-Simons number.
Kinetic theory descriptions of an evolving quark gluon plasma find
variations in the Chern-Simons number \cite{Arnold} too. Ultimately
one hopes to understand quantitatively how the glasma evolves into the
quark gluon plasma. This then could give us a more reliable estimate
of the amount of chirality change for very early times.

\section{The Chiral Magnetic Effect}
In order to observe $\mathcal{P}$ and $\mathcal{CP}$ violation on an
event-by-event basis in heavy ion collisions we have to know how to
distinguish left- from right-handed quarks.  Measuring directly the
helicity of the quarks is impossible since one only measures hadrons
in the detectors.  The solution is polarization, which, as we will see
in the next section, can be generated by the (electromagnetic) magnetic
field created by the colliding ions.

\begin{figure}[th]
\begin{center}
\includegraphics[width=8cm]{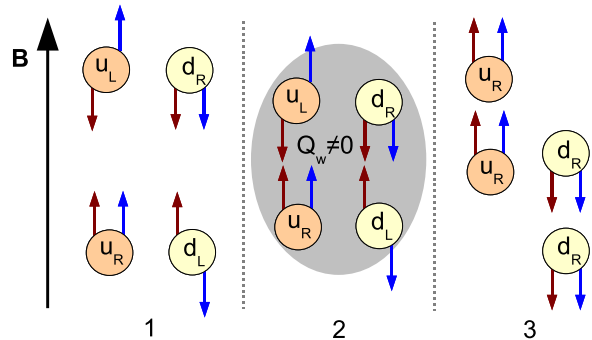}
\end{center}
\caption{Illustration of the Chiral Magnetic Effect in a very
  large homogeneous magnetic field.   The red arrows denote the direction
  of momentum, the blue arrows -- the spin of the quarks. 
  (1) The magnetic field polarizes the quarks. (2) The quarks
  interact with gluon fields which have $\qw = -1$ in this case. (3)
  The chirality change induces an electromagnetic current along
  the direction of the magnetic field.
 }
\label{fig:chargesep}
\end{figure}

Let us for a moment assume that we have a homogeneous magnetic field
$B$ pointing in the $z$-direction.  Let us furthermore assume that the
magnitude of $B$ is much larger than the square of the typical
momentum of the quarks. In that case the quarks will be fully
polarized along the magnetic field. This means that the spins of the
quarks align along the direction of the magnetic field depending on
their electric charge.  But since the quarks are massless, the momenta
of the quarks will also align along the direction of the magnetic
field.  We have illustrated this situation in Fig.\
\ref{fig:chargesep}.  The left- and
right-handed quarks will now move in opposite directions, hence we can
distinguish them.

After the quarks have interacted with the gluon fields a difference
between the number of left- and right-handed quarks is generated which
is equal to $2 \qw$. Because left- and right-handed quarks
move in opposite directions, an electromagnetic current is
set up along the magnetic field.  This is the Chiral Magnetic Effect.

If the electromagnetic current flows in a finite volume, a charge
difference of magnitude $Q = 2 \qw \sum_f \vert q_f \vert$ will be
generated between the upper and lower hemisphere. Here $q_f$ denotes
the charge of a quark with flavor $f$. This charge difference
is the same if antiquarks change their chirality or when 
quark-antiquark pairs are produced from a gluon field
with a winding number.

If the magnetic field is moderate, quarks with high momentum
will be less polarized.  Let us denote the degree of polarization of a
quark as $\gamma(q_f)$. The expectation value of the induced charge
difference will now become $Q = 2 \qw \sum_f \vert q_f \vert
\gamma(q_f)$.  Only quarks which have momenta smaller than the
inverse size $1/\rho$ of the gauge field configurations with winding
number $\qw$ will interact, and hence change their helicity. Therefore
only the polarization of quarks with momentum smaller than $1/\rho$ is
relevant. In \cite{KMW} we found that the polarization can be
estimated by the following formula $\gamma(q_f) \approx 2 \vert q_f e
B \vert \rho^2$.

The typical size of a sphaleron is bounded by the
chromomagnetic screening length of the quark gluon plasma which 
is $\rho \sim 1/(\alpha_S T)$. Hence in order to get
reasonable polarization the magnetic field has to be
of order $\alpha_s^2 T^2$, which is $10^3 - 10^4\;\mathrm{MeV^2}$.
This corresponds to $10^{13}-10^{14}\; \mathrm{T}$. In the next section
we will see that such huge fields are created in heavy ion
collisions.

\section{The implications for heavy ion collisions}
\begin{figure}[t]
\begin{center}
\includegraphics[width=12cm]{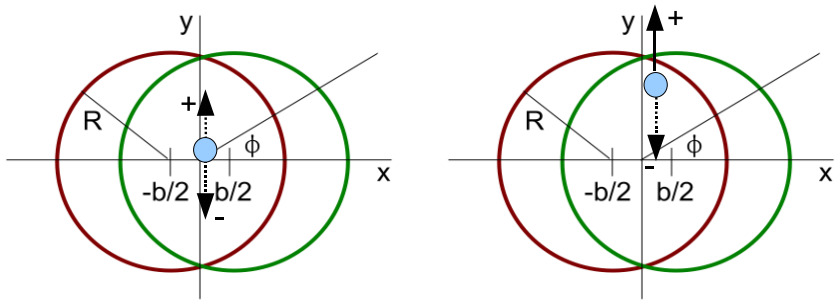}
\end{center}
\caption{Two cross-sectional views of a non-central heavy ion collision along
  the beam-axis ($z$-axis). The plane $y=0$ is called the reaction
  plane.  The region in which the two nuclei overlap contains the
  participants, the regions in which they do not contain the
  spectators. Two examples of sphaleron transitions (indicated with a
  small circle) are depicted.  The contribution to the separation of
  charge of the transitions near the center (left-hand side) is
  suppressed with respect to transitions near the surface (right-hand
  side) due to screening.  }
\label{fig:collision}
\end{figure}

\begin{figure}[htb]
\begin{center}
\begin{tabular}{cc}
\includegraphics[width=6cm]{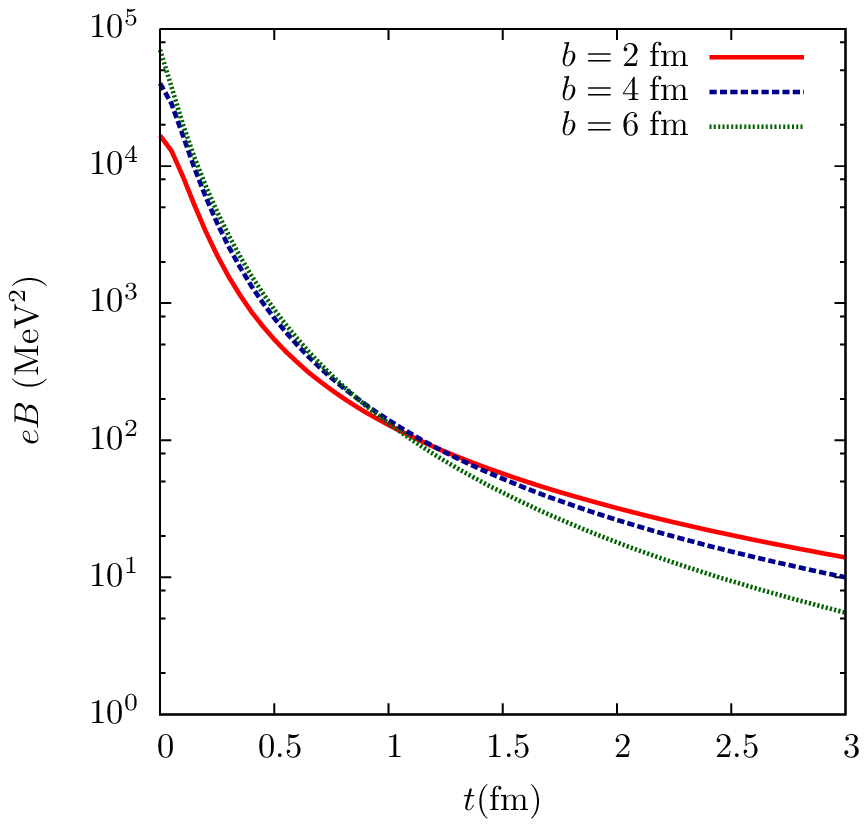}
&
\includegraphics[width=6cm]{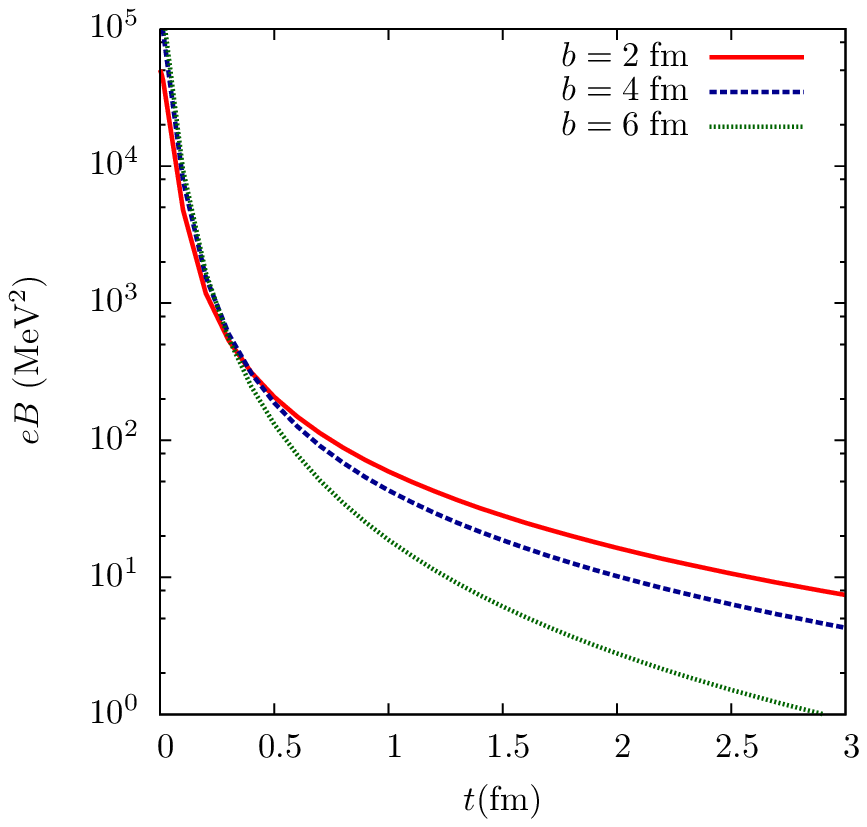}
\end{tabular}
\end{center}
\caption{Magnetic field at the center of a Gold-Gold
collision, for different impact
parameters. In left plot the center of mass energy is 
$62\;\mathrm{GeV}$ per nucleon pair,
in the right plot it is 
$200\;\mathrm{GeV}$ per nucleon pair.}
\label{fig:magfield}
\end{figure}

For reference we have displayed a cross-sectional view of a heavy ion
collision in Fig.~\ref{fig:collision}.  If two heavy ions collide
enormous magnetic fields are generated in the direction of angular
momentum (the $y$-direction). We have computed the magnetic field
generated by the spectators and participants using a classical
calculation which included the effect of baryon stopping in the
participant region \cite{KMW}. The results are displayed in
Fig.~\ref{fig:magfield}. The magnetic field is so large because of the
large charge of the nuclei and the short distances and time scales
involved. The magnetic field is capable of polarizing quarks to a certain
degree just after the collision.

Because the quarks will be polarized along the $y$-axis, the Chiral
Magnetic Effect will separate quarks along this axis. Now imagine a
sphaleron transition taking place in the center of the participant
region (left-hand side of Fig.~\ref{fig:collision}). The
quarks that are being separated would still experience
interactions, so that presumably their momenta get randomized.  If the
sphaleron transition takes place at near the surface of the
participant region  (right-hand side of Fig.~\ref{fig:collision}),  
a quark with one kind of charge is able to
hadronize without experience many interactions, while the quark with opposite
charge still has to travel through the hot medium.  Since charge is
conserved in hadronization, asymmetries in charge generated by quarks,
will end up into asymmetries in charged hadrons.

Let us now define $\Delta_\pm$ to be the difference between the amount of
positive / negative charge above and below the reaction plane.  We have
argued in Ref.~\cite{KMW} that every time a topological
charge changing transition is made, $\Delta_\pm$ is modified as follows
\begin{equation}
 \Delta_+ \rightarrow \Delta_+ \pm \sum_f \vert q_f \vert \gamma(q_f)
 \xi_\pm(x_\perp),
\;\;\;\;\;
 \Delta_- \rightarrow \Delta_- \mp \sum_f \vert q_f \vert \gamma(q_f)
 \xi_\mp(x_\perp) .
\label{eq:deltaq}
\end{equation}
Here $\xi_\pm(x_\perp)$ is a phenomenological screening function.  By
folding rate of sphaleron transitions with the square of the changes
from Eq.~(\ref{eq:deltaq}) and integrating over time and the volume of
the participant region it is possible to compute the variance of
$\Delta_\pm$ and the correlation $\langle \Delta_+ \Delta_- \rangle$
\cite{KMW}. Since the magnetic field decreases rapidly as a function
of time, the main contribution to the correlations comes from early
times. 

In \cite{KMW} we have studied the magnitude of $a_{ij} \sim \langle
\Delta_i \Delta_j \rangle / (N_i N_j)$. Here $i, j = \pm$ and $N_\pm$
denotes the total number of particles of a particular charge. We have
estimated that $a_{ij}$ for large impact parameters can be of order
$10^{-4}$, with order of magnitude uncertainties. This means that the
typical expected asymmetries could be of order $1 \%$. We expect that
$a_{ij}$ will increase as a function of impact parameter, since for
larger impact parameters the magnetic field is larger and the
screening is relatively less important. Since $a_{+-}$ is affected
more by the screening effect, $\vert a_{+-}/a_{++} \vert$ is expected
to grow as a function of impact parameter.

The correlators $a_{ij}$ can be measured
using the method proposed by Voloshin \cite{Voloshin2004}. In order
to obtain the correlators one has to average over many events. 
The observables itself are not $\mathcal{P}$- and $\mathcal{CP}$-odd,
in a sense one is measuring the absolute value of the dipole moment.
Preliminary data from the STAR collaboration which is presented in
Ref.~\cite{IVS} show hints that such correlations might exist.

\section{Conclusions}
The Chiral Magnetic Effect is a signature for $\mathcal{P}$-
and $\mathcal{CP}$-violation on an event-by-event basis.
It will be direct evidence for the existence of topologically
nontrivial configurations of gauge fields, and therefore
complement searches for instantons in scattering experiments \cite{DIS}.

Observation of the Chiral Magnetic Effect will be evidence for the
existence of a quark gluon plasma, a phase in which matter is deconfined
and chiral symmetry is restored.  The reason is that if the quarks are
confined, we cannot separate them.  Moreover, if chiral symmetry is
broken, quarks become effectively massive, which removes the necessary
correlation between the spin and momentum of the quarks.

In order to obtain our results, we had to make some approximations. 
It would be desirable to improve our results to more obtain
accurate predictions for the absolute magnitude and
the dependence of the Chiral Magnetic Effect
on the kind of nucleus, the beam energy and the impact parameter.
Nevertheless, the Chiral Magnetic Effect is a very natural
consequence of QCD in the presence of a strong magnetic field.
It is therefore conceivable that it might
be observed in heavy ion collisions.

Let us finally point out that the Chiral Magnetic Effect is in some
sense similar to Electroweak Baryogenesis in the early universe. There
a $\mathcal{C}$- and $\mathcal{CP}$-violating sphaleron transition
induces via the axial anomaly a difference in the number of baryon
plus lepton number \cite{hooft}.  In some scenarios this then could
lead to the observed baryon asymmetry of the universe
\cite{baryonasym}. If however large magnetic fields were present in
the early universe during the QCD phase transition, the Chiral
Magnetic Effect itself might even have implications for cosmology
\cite{FZ}.

\section*{Acknowledgments}
I would like to thank the organizers of Quark Matter 2008 for the
wonderful conference and the opportunity to present this work.  
I am very grateful to Dmitri Kharzeev and Larry McLerran 
for discussions and the collaboration which led
to the paper \cite{KMW} on which this work is based.
I would like to thank Vasily Dzordzhadze, Rob Pisarski, Jianwei Qiu,
Ilya Selyuzhenkov, Yannis Semertzidis and Sergei Voloshin for
useful discussions.  This manuscript has been authored under Contract
No.~\#DE-AC02-98CH10886 with the U.S.\ Department of Energy.

\end{document}